\documentclass[twocolumn,showpacs,preprintnumbers,amsmath,amssymb,prl]{revtex4}

\usepackage{graphicx}
\usepackage{bm}

\begin{document}

\title{Water-like Anomalies of Core-Softened Fluids:
Dependence on the Trajectories in ($P\rho T$) Space}

\author{Yu. D. Fomin}
\affiliation{Institute for High Pressure Physics, Russian Academy
of Sciences, Troitsk 142190, Moscow Region, Russia}

\author{V. N. Ryzhov}
\affiliation{Institute for High Pressure Physics, Russian Academy
of Sciences, Troitsk 142190, Moscow Region, Russia}

\date{\today}

\begin{abstract}
In the present article we carry out a molecular dynamics study of
the core-softened system and show that the existence of the
water-like anomalies in this system depends on the trajectory in
$P-\rho-T$ space along which the behavior of the system is
studied. For example, diffusion and structural anomalies are
visible along isotherms, but disappears along the isochores and
isobars, while density anomaly exists along isochors.  We analyze
the applicability of the Rosenfeld entropy scaling relations to
this system in the regions with the water-like anomalies. It is
shown that the validity of the of Rosenfeld scaling relation for
the diffusion coefficient also depends on the trajectory in the
$P-\rho-T$ space along which the kinetic coefficients and the
excess entropy are calculated.
\end{abstract}

 \maketitle

\section{I. Introduction}

It is well known that some liquids (for example, water, silica,
silicon, carbon, and phosphorus) show an anomalous behavior in the
vicinity of their freezing lines \cite{deben2003,bul2002,
angel2004,book,book1,deben2001,netz,stanley1,ad1,ad2,ad3,ad4,ad5,
ad6,ad7,ad8,errington,errington2}. The water phase diagrams have
regions where a thermal expansion coefficient is negative (density
anomaly), self-diffusivity increases upon compression (diffusion
anomaly), and the structural order of the system decreases with
increasing pressure (structural anomaly) \cite{deben2001,netz}.
The regions where these anomalies take place form nested domains
in the density-temperature \cite{deben2001} (or
pressure-temperature \cite{netz}) planes: the density anomaly
region is inside the diffusion anomaly domain, and both of these
anomalous regions are inside a broader structurally anomalous
region. It is reasonable to relate this kind of behavior to the
orientational anisotropy of the potentials, however, a number of
studies demonstrate waterlike anomalies in fluids that interact
through spherically symmetric potentials
\cite{8,9,10,11,12,13,14,15,16,17,18,19,20,bar2,21,22,23,24,25,26,
barb2008-1,barb2008,FFGRS2008,RS2002,RS2003,FRT2006,GFFR2009,fr1,fr2}.

As it was mentioned above, three types of anomalies are the most
discussed in literature with respect to the core-softened fluids -
diffusion anomaly, density anomaly and structural anomaly. Here we
briefly discuss these anomalies.

If we consider a simple liquid (for, example, Lennard-Jones
liquid), and trace the diffusion along an isotherm we find that
the diffusion decreases under densification. This observation is
intuitively clear - if density increases the free volume decreases
and the particles have less freedom to move. However, some
substances have a region in density - temperature plane where
diffusion grows under densification. This is called anomalous
diffusion region which reflects the contradiction of this behavior
with the free volume picture described above. This means that
diffusion anomaly involves more complex mechanisms which will be
discussed below.

The second anomaly mentioned above is density anomaly. It means
that density increases upon heating or that the thermal expansion
coefficient becomes negative. Using the thermodynamic relation
$\left(\partial P/\partial T\right)_V=\alpha_P/K_T$, where
$\alpha_P$ is a thermal expansion coefficient and $K_T$ is the
isothermal compressibility and taking into account that $K_T$ is
always positive and finite for systems in equilibrium not at a
critical point, we conclude that density anomaly corresponds to
minimum of the pressure dependence on temperature along an
isochor. This is the most convenient indicator of density anomaly
in computer simulation.

The last anomaly we discuss here is structural anomaly. Initially
this anomaly was introduced via order parameters characterizing
the local order in liquid. However, later on the local order was
also related to excess entropy of the liquid which is defined as
the difference between the entropy and the ideal gas entropy at
the same $(\rho,T)$ point: $S_{ex}=S-S_{id}$. In normal liquid
excess entropy is monotonically decaying function of density along
an isotherm while in anomalous liquids it demonstrates increasing
in some region. This allows to define the boundaries of structural
anomaly at given temperature as minimum and maximum of excess
entropy.

The problem of anomalous behavior of core-softened fluids was
widely discussed in literature (see, for example, the recent
review \cite{fr1}). It was shown that for some systems the
anomalies take place while for others does not. In this respect
the question of criteria of anomalous behavior appearance remains
the central one. However, another important point is still lacking
in the literature - the behavior of anomalies along different
thermodynamic trajectories. Here we call as "trajectory" a set of
points belonging to some path in $(P,\rho,T)$ space. For example,
the set of points belonging to the same isotherm we call as
"isothermal trajectory" or shortly isotherm.

In our previous work we showed \cite{werostr} that anomalies can
exist along some trajectories while along others the liquid
behaves as a simple one. Taking into account this result, it is
interesting to study the behavior of the quantities demonstrating
anomalies along the different physically significant trajectories
(isotherms, isochors, isobars and adiabats). This investigation
will allow to get deeper understanding of the relations between
anomalous behavior and thermodynamic parameters of the system
which spread light on the connection between thermodynamic,
structural and dynamic properties of liquids.

\section{II. System and Methods}

The simplest form of core-softened potential is the so called
Repulsive Step Potential which is defined as following:
\begin{equation}
\Phi (r)=\left\{
\begin{array}{lll}
\infty , & r\leq d \\
\varepsilon , & d <r\leq \sigma  \\
0, & r>\sigma%
\end{array}%
\right.  \label{1}
\end{equation}
where $d$ is the diameter of the hard core, $\sigma$ is the width
of the repulsive step,  and  $\varepsilon$ its height. In the
low-temperature limit $\tilde{T}\equiv k_BT/\varepsilon<<1$  the
system reduces to a hard-sphere systems with hard-sphere diameter
$\sigma$, whilst in the limit $\tilde{T}>>1$ the system reduces to
a hard-sphere model with a hard-sphere diameter $d$. For this
reason, melting at high and low temperatures follows simply from
the hard-sphere melting curve $P=cT/\sigma'^3$, where $c \approx
12$ and $\sigma'$ is the relevant hard-sphere diameter ($\sigma$
and $d$, respectively). A changeover from the low-$T$ to high-$T$
melting behavior should occur for $\tilde{T} ={\mathcal O}(1)$.
The precise form of the phase diagram depends on the ratio
$s\equiv \sigma/d$. For large enough values of $s$ one should
expect to observe in the resulting melting curve a maximum that
should disappear as $s\rightarrow 1$. The phase behavior in the
crossover region may be very complex, as  shown
in~\cite{FFGRS2008,GFFR2009}.

In the present simulations we have used a smoothed version of the
repulsive step potential (Eq.~(\ref{1})), which has the form:
\begin{equation}
\Phi (r)=
\left(\frac{d}{r}\right)^{n}+\frac{1}{2}\varepsilon\left(1-\tanh\left(k_0
\left(r-\sigma_s \right)\right)\right) \label{2}
\end{equation}
where $n=14,k_0=10$. We have considered $\sigma_s= 1.35$. Here and
below we refer to this potential as to smooth Repulsive Shoulder
System (RSS).

In the remainder of this paper we use the dimensionless
quantities: $\tilde{{\bf r}}\equiv {\bf r}/d$, $\tilde{P}\equiv P
d^{3}/\varepsilon ,$ $\tilde{V}\equiv V/N d^{3}\equiv
1/\tilde{\rho}$. As we will only use these reduced variables, we
omit the tildes.

In Refs.~\cite{FFGRS2008,GFFR2009},  phase diagrams of
repulsive-step models were reported for $\sigma_s=1.15, 1.35,
1.55, 1.8$. To determine the phase diagram at non-zero
temperature, we performed NVT MD simulations combined with
free-energy calculations. For equilibration rescaling of the
velocities was used, for sampling we used NVE ensemble monitoring
the stability of the temperature. In all cases, periodic boundary
conditions were used. The number of particles varied between 250,
500 and 864. No system-size dependence of the results was
observed. The system was equilibrated for $5\times 10^6$ MD time
steps. Data were subsequently collected during $3\times 10^6\delta
t$ where the time step $\delta t=5\times 10^{-5}$.

In order to map out the phase diagram of the system, we computed
its Helmholtz free energy using thermodynamic integration: the
free energy of the liquid phase was computed via thermodynamic
integration from the dilute gas limit~\cite{book_fs}, and the free
energy of the solid phase was computed by thermodynamic
integration to an Einstein crystal~\cite{book_fs,fladd}.  In the
MC simulations of solid phases, data were collected during $5
\times 10^4$ cycles after equilibration. To improve the statistics
(and to check for internal consistency) the free energy of the
solid was computed at many dozens of different state-points and
fitted to multinomial function. The fitting function we used is
$a_{p,q}T^pV^q$, where $T$ and $V=1/\rho$ are the temperature and
specific volume and powers $p$ and $q$ are connected through $p+q
\le N$. The value N we used for the most of calculations is 5. For
the low-density FCC phase  N was taken equal to 4, since we had
less data points. The transition points were determined using a
double-tangent construction.

The region where thermodynamic anomalies are expected is situated
close to the region where the system undergoes structural arrest.
As a consequence, proper sampling of the phase space can be
problematic. To overcome this problem we used  parallel
tempering~\cite{book_fs}. Instead of simulating a single system,
we consider $n$ systems, each running in the NVT ensemble at a
different temperature. Systems at high temperatures go easily over
potential barriers and systems at low temperatures sample the
local free energy minima. The idea of parallel tempering is to
attempt Monte Carlo temperature swaps  between the systems at
different temperatures. If the low and high temperatures are far
apart, the probability to accept such a swap move is quite low.
For this reason, we use a range of 'intermediate' temperatures.
During a parallel tempering simulation, we generate equilibrium
configurations for the system at all the temperatures in the
simulation. In most cases, we used 8 temperatures and tried to
swap them 40 times. This simulation took almost 24 hours running
on 8 processors in parallel at the Joint Supercomputing Center of
Russian Academy of Sciences.

\begin{figure}
\includegraphics[width=7cm, height=7cm]{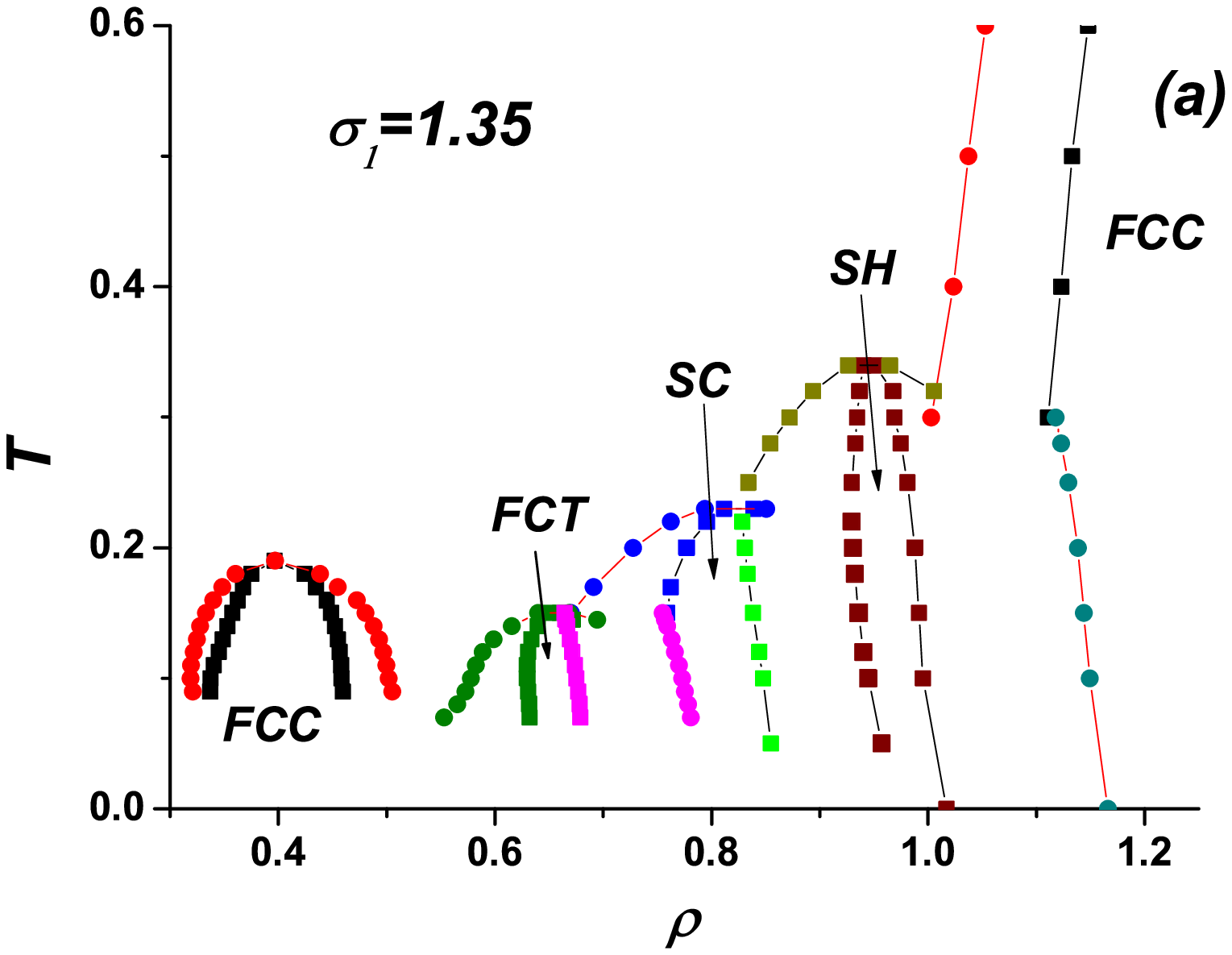}

\includegraphics[width=7cm, height=7cm]{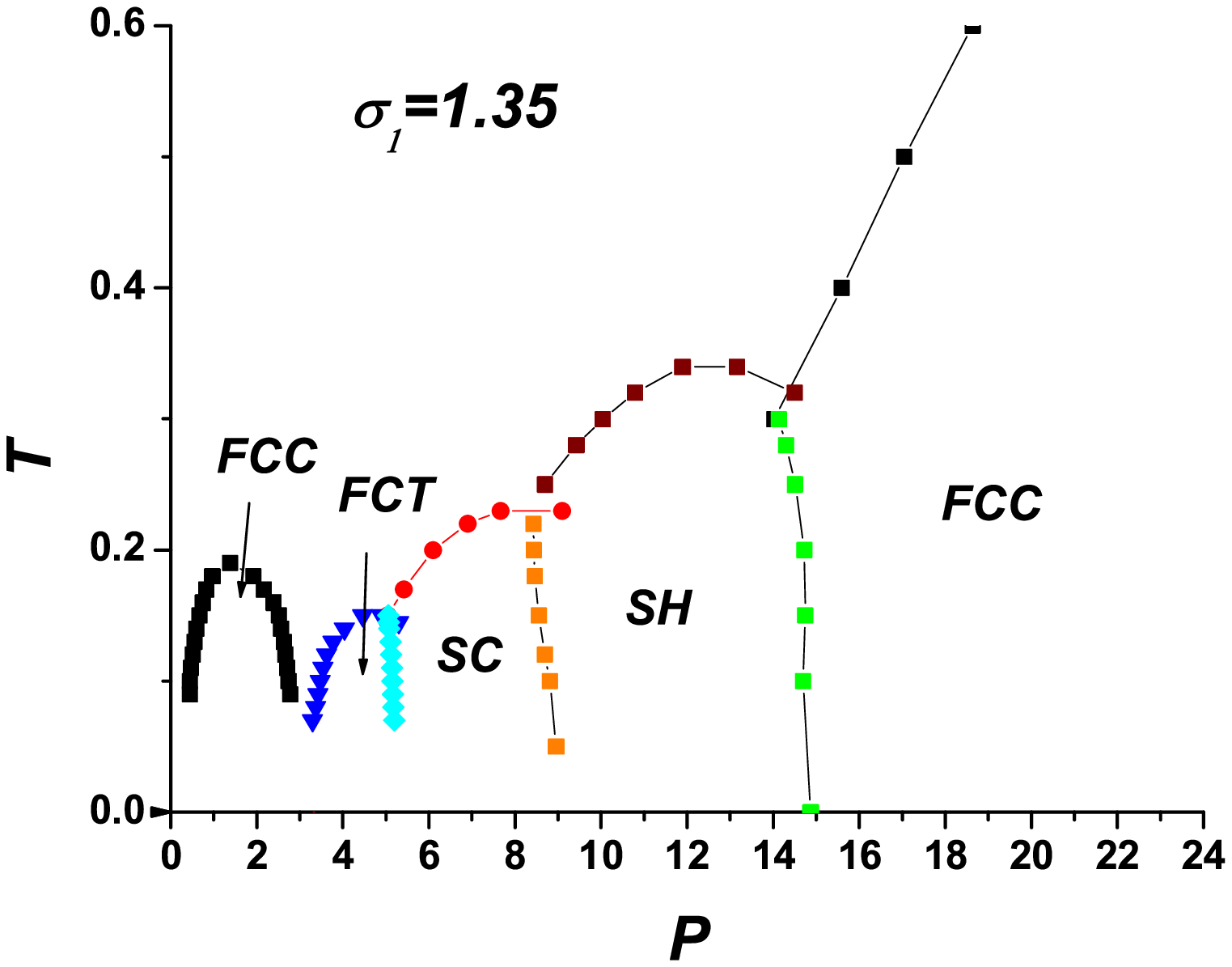}%

\caption{\label{fig:fig1a} (Color online). Phase diagram of the
system of particles interacting through the potential (2) with
$\sigma_s= 1.35$ in $\rho-T$ and $P-T$ planes.  }
\end{figure}

Fig.~\ref{fig:fig1a} shows the phase diagrams that we obtain from
the free-energy calculations for $\sigma_s=1.35$. In fact, the
phase diagrams for $\sigma_s=1.15, 1.35, 1.55, 1.8$ were already
reported in Refs.~\cite{FFGRS2008,GFFR2009}. We show these phase
diagrams here too because they provide the ``landscape'' in which
possible ``water'' anomalies should be positioned.

Fig.~\ref{fig:fig1a}(a) shows the phase diagram of the system with
$\sigma_s=1.35$ in the $\rho-T$ plane. There is a clear maximum in
the melting curve at low densities. The phase diagram consists of
two isostructural FCC domains corresponding to close packing of
the small and large spheres separated by a sequence of structural
phase transitions. This phase diagram was discussed in detail in
our previous publication \cite{FFGRS2008,GFFR2009}. It is
important to note that there is a region of the phase diagram
where we have not found any stable crystal phase. The results of
Ref.~ \cite{FFGRS2008} suggest that a glass transition occurs in
this region with $T_g=0.079$ at $\rho=0.53$. The apparent
glass-transition temperature is above the melting point of the
low-density FCC and FCT phases. If, indeed, no other crystalline
phases are stable in this region, the ``glassy'' phase that we
observe would be thermodynamically stable. This is rather unusual
for one-component liquids. In simulations, glassy behavior is
usually observed in metastable mixtures, where crystal nucleation
is kinetically suppressed. One could argue that, in the glassy
region, the present system behaves like a ``quasi-binary'' mixture
of spheres with diameters $d$ and $\sigma_s$ and that the
freezing-point depression is analogous to that expected in a
binary system with a eutectic point: there are some values of the
diameter ratio such that crystalline structures are strongly
unfavorable and the glass phase could then be stable even at very
low temperatures. The glassy behavior in the reentrant liquid
disappears at higher temperatures.

\section{III. Results and discussion}

\subsection{Diffusion anomaly}

The diffusion anomaly of the RSS was discussed in several our
previous articles. In the Refs. \cite{FFGRS2008} and
\cite{GFFR2009} we showed that the diffusion anomaly takes place
at the shoulder width $\sigma_1=1.35$ while in the work
\cite{weros} the breakdown of the Rosenfeld scaling for this
system was demonstrated. Further discussion of the Rosenfeld
scaling was reported in the work \cite{werostr}. The main idea of
this paper is that the anomalous diffusion behavior present along
low-temperature isotherms while along isochors diffusion
coefficient is monotonous. As a result, Rosenfeld scaling is valid
along isochors and high-temperature isotherms, however, it breaks
down for the isotherms with anomalies. It means that appearance or
not of anomalies at some trajectory can cause physically different
behavior of the system. As a result, following different
trajectories, we can or can not observe some effects. It is of
particular importance since experimental works deal mostly with
isobars and isotherms while theoretical studies - with isochors
and isotherms. Taking this into account, one can expect that some
of the effects observed along isobars are not visible along
isotherms and isochors which makes us confused while comparing
experimental results with theoretical predictions. Here we extend
the study of anomalies to four different physically meaningful
trajectories: isotherms, isochors, isobars and adiabats.

\bigskip

\begin{figure}
\includegraphics[width=7cm, height=7cm]{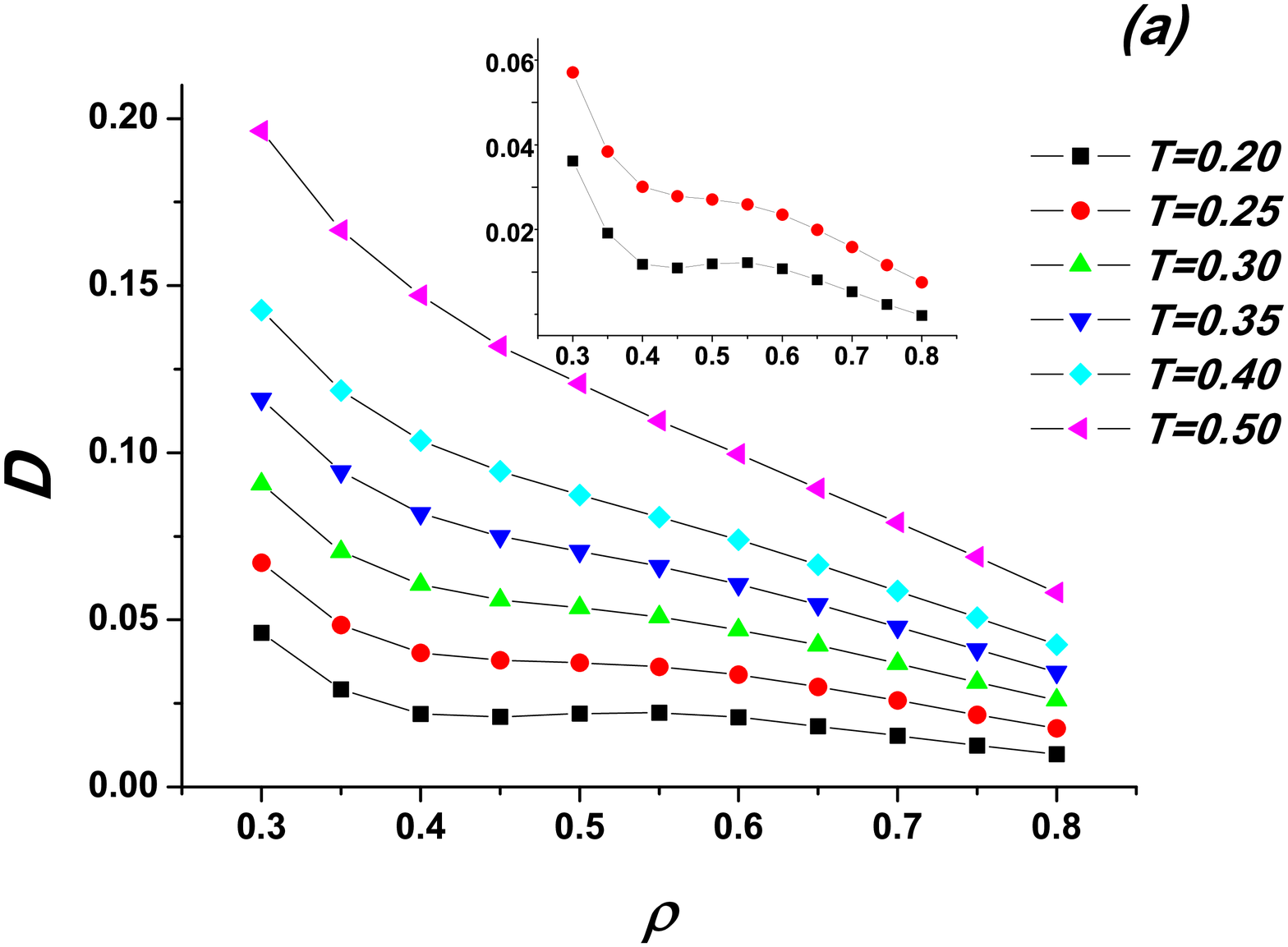}%

\includegraphics[width=7cm, height=7cm]{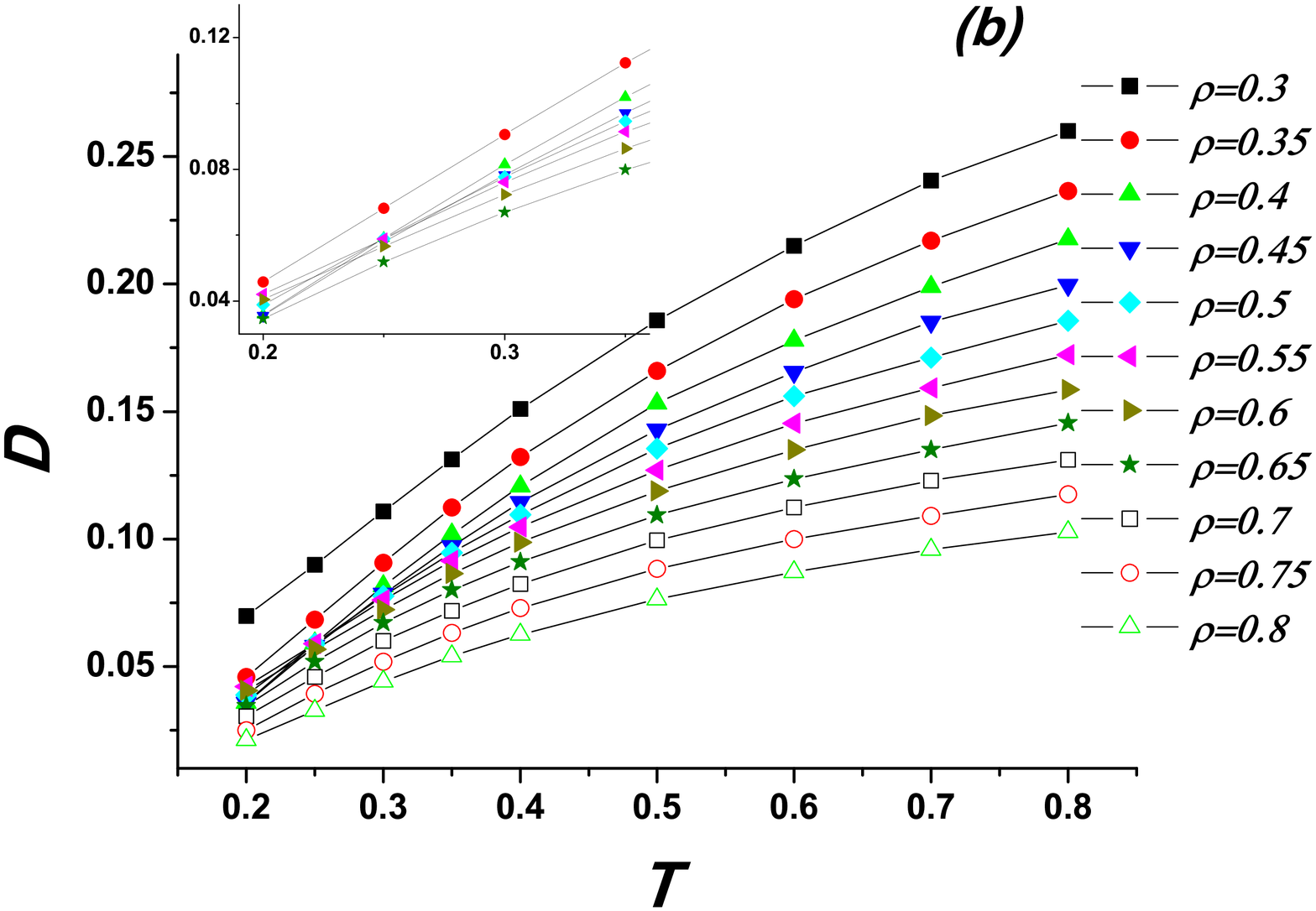}%
\caption{\label{fig:fig1} (Color online). Diffusion coefficient of
the RSS system along (a) isotherms and (b) isochors. The insets
show (a) the low temperature isotherms and (b) low temperature
region of some isochors. }
\end{figure}

Fig.~\ref{fig:fig1}(a) shows the diffusion coefficient of RSS
system along a set of isotherms for $\sigma_1=1.35$. One can see
that the diffusion coefficient demonstrates anomalous behavior for
the temperatures below $T=0.25$. However, if we look at the
Fig.~\ref{fig:fig1}(b) where the same diffusion coefficient data
are arranged along isochors we do not observe anomalies - the
diffusion is a monotonous function of temperature along isochors.
However, some of the isochors cross. One can see from the inset
that the cross of the isochors corresponds to the densities
between $\rho=0.4$ and $\rho=0.6$. Comparing it to the isotherms
we see that this is the region of anomalous diffusion. It means
that even if we do not see nonmonotonous behavior of diffusion
along isochors we can identify the presence of anomaly from
crossing of the isochors. However, this method seems to be
technically more difficult since we need to measure many points
belonging to different isochors rather then one isotherm.

\begin{figure}
\includegraphics[width=7cm, height=7cm]{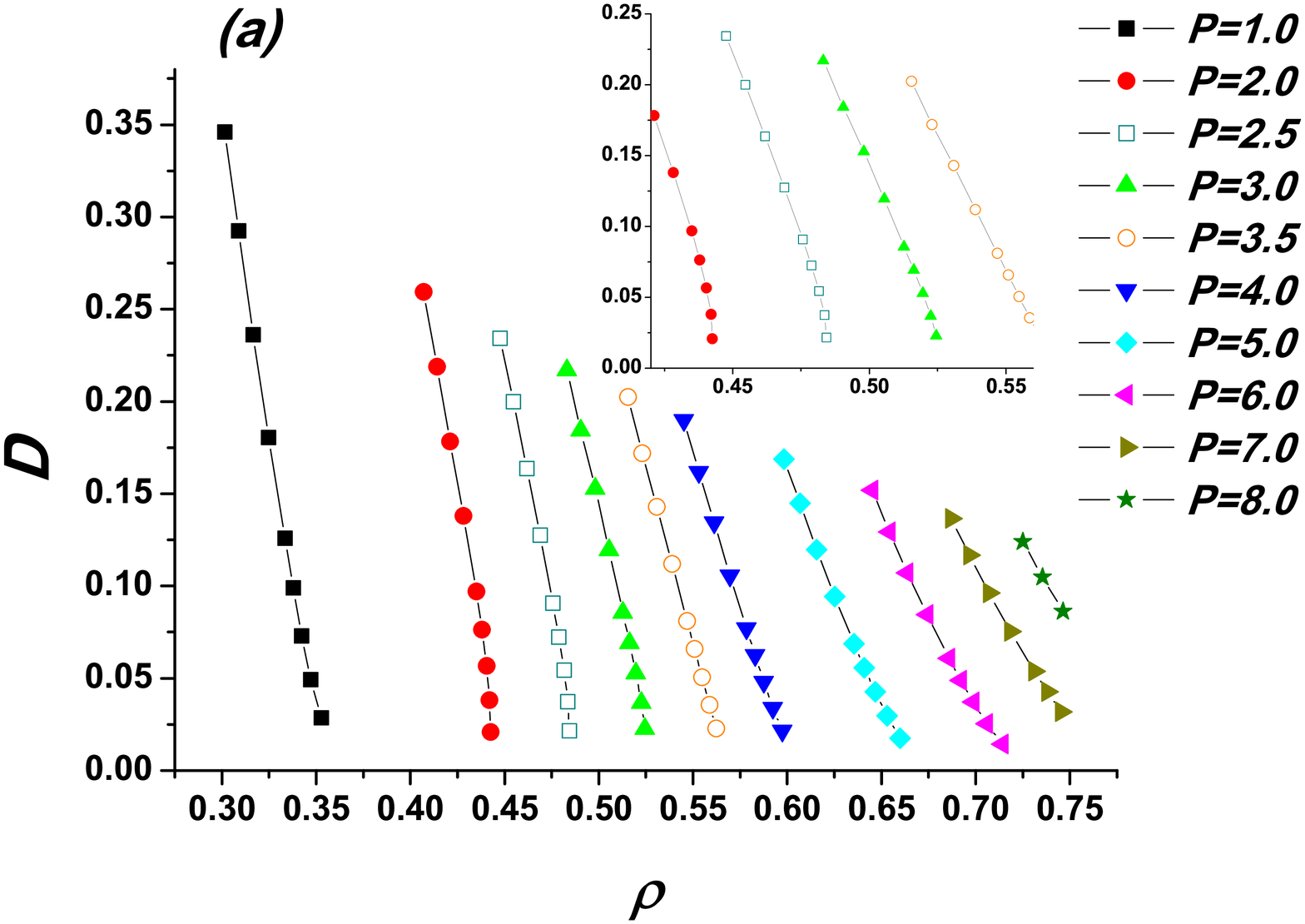}%

\includegraphics[width=7cm, height=7cm]{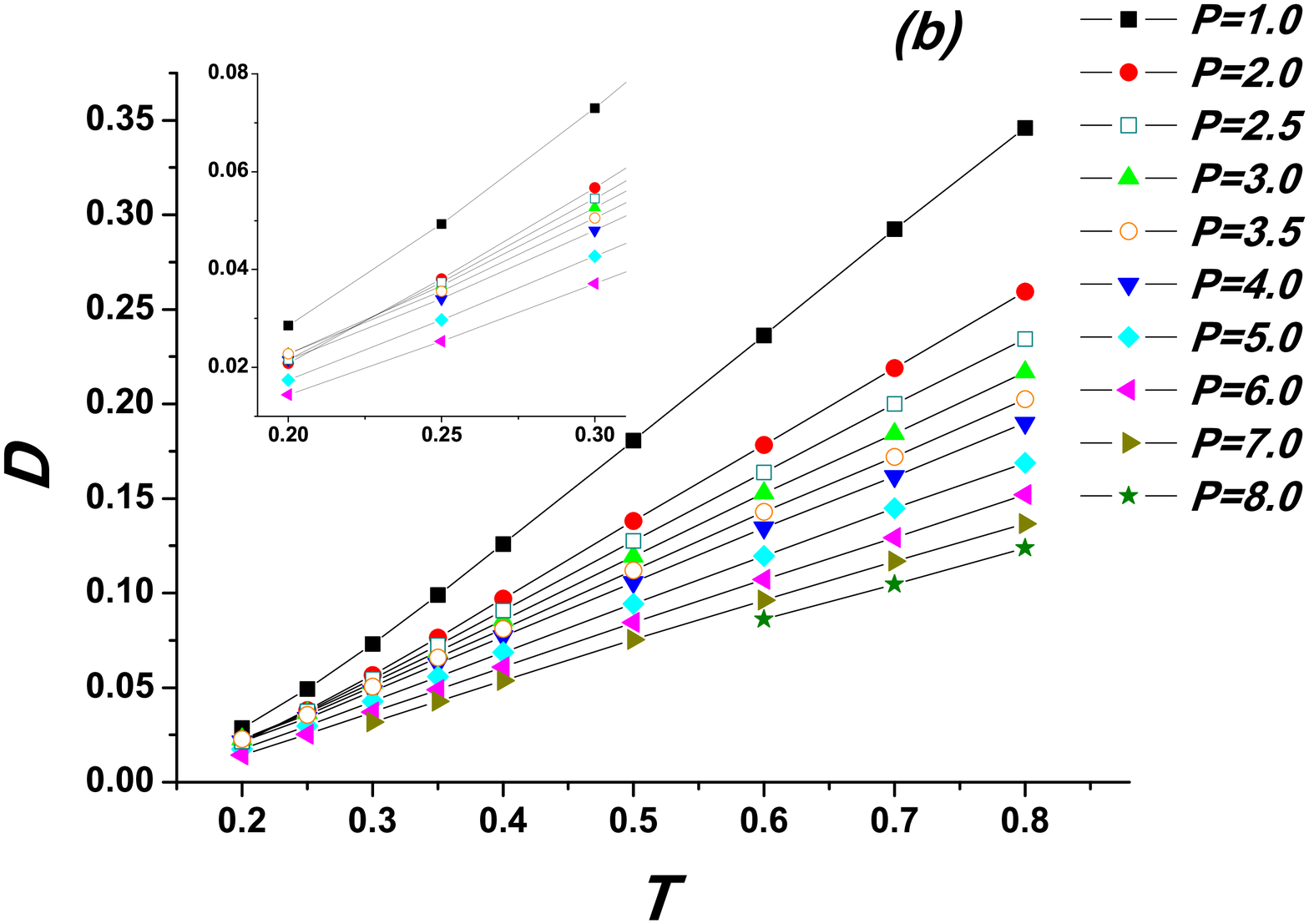}%
\caption{\label{fig:fig2} (Color online). Diffusion coefficient of
the RSS system along isobars as a function of (a) density and (b)
temperature. The insets show anomalous regions in the
corresponding coordinates.}
\end{figure}

Isothermal and isochoric behavior of diffusion in anomalous region
have already been discussed in our previous publication
\cite{werostr} and we give these plots here for the sake of
completeness. Fig.~\ref{fig:fig2}(a) shows the diffusion
coefficient along a set of isobars as a function of density. The
diffusion coefficient is again monotonous. The slope of the curves
is always negative. However, as it can be seen from the inset, the
slope approaches infinity at low diffusions at pressures $P=2.0$
and $P=2.5$. This corresponds to densities $0.45-0.55$ inside the
anomalous region. One can imagine that if we lower the
temperatures along these isobars we can observe change of the
slope to positive one, however, we do not have data for these
temperatures.

Fig.~\ref{fig:fig2}(b) shows the diffusion coefficient along
isobars as a function of temperature. The situation is analogous
to the case of isochors: the curves are monotonous, however, they
intersect at low temperatures, corresponding to anomalous region
(see the inset in Fig.~\ref{fig:fig2}(b)). It means that if we
have the diffusion coefficient along isobars we can identify the
presence of anomalies by monitoring the intersections of the
curves. However, by the reasons discussed above this method is not
practically convenient.

The last physically meaningful trajectory considered in the
present work is the adiabat. This trajectory is defined as
constant entropy curve. The entropy is calculated as following. We
compute excess free energy by integrating the equation of states:
$\frac{F_{ex}}{Nk_BT}=\frac{F-F_{id}}{Nk_BT}=\frac{1}{k_BT}
\int_0^{\rho} \frac{P(\rho ')-\rho ' k_BT}{\rho '^2} d\rho'$. The
excess entropy can be computed via $S_{ex}=\frac{U-F_{ex}}{N
k_BT}$. The total entropy is $S=S_{ex}+S_{id}$, where the ideal
gas entropy is
$\frac{S_{id}}{Nk_B}=\frac{3}{2}\ln(T)-\ln(\rho)+\ln(\frac{(2 \pi
mk_B)^{3/2}e^{5/2}}{h^3})$. The last term in this expression is
constant and is not accounted in our calculations.

\begin{figure}
\includegraphics[width=7cm, height=7cm]{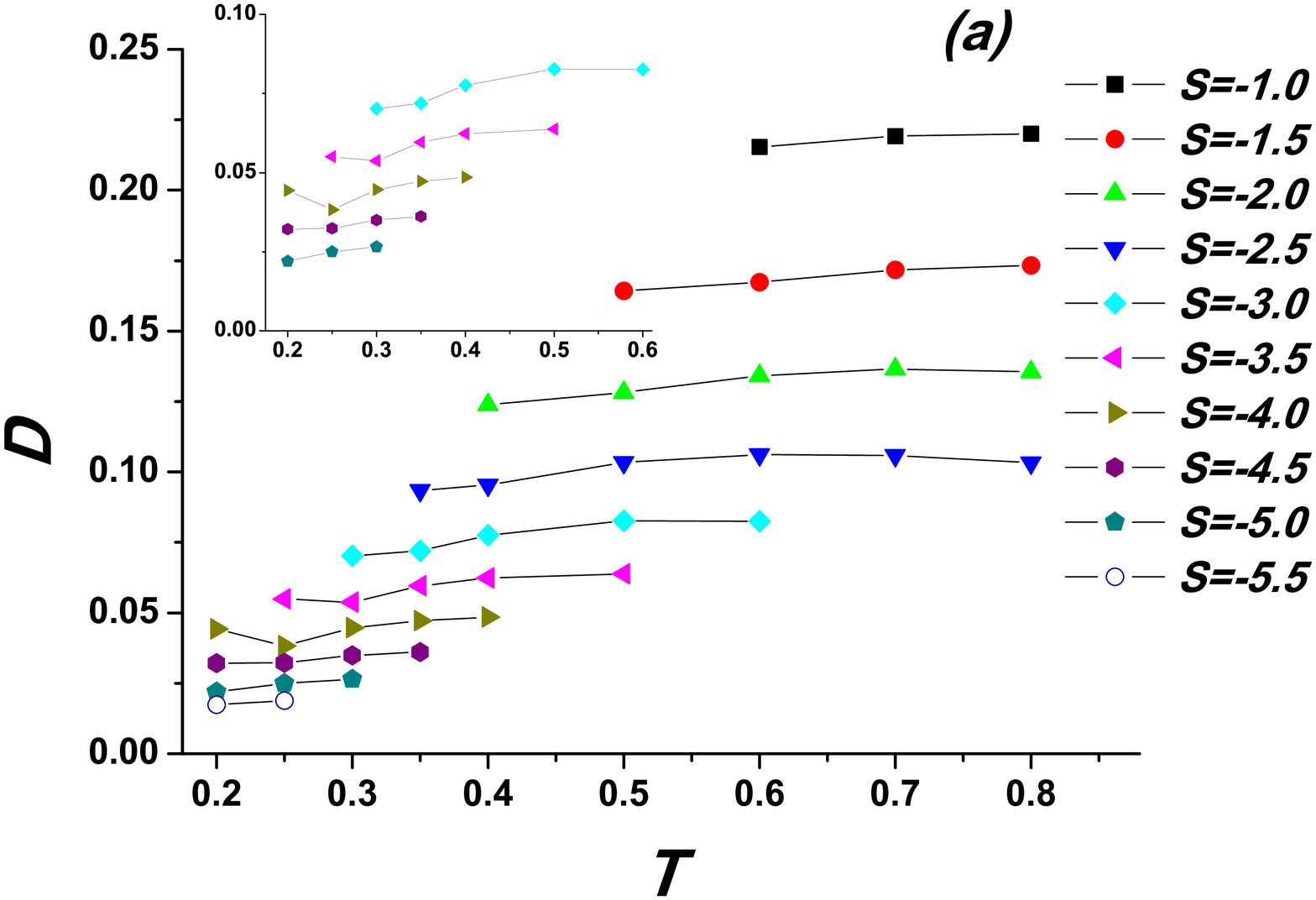}%

\includegraphics[width=7cm, height=7cm]{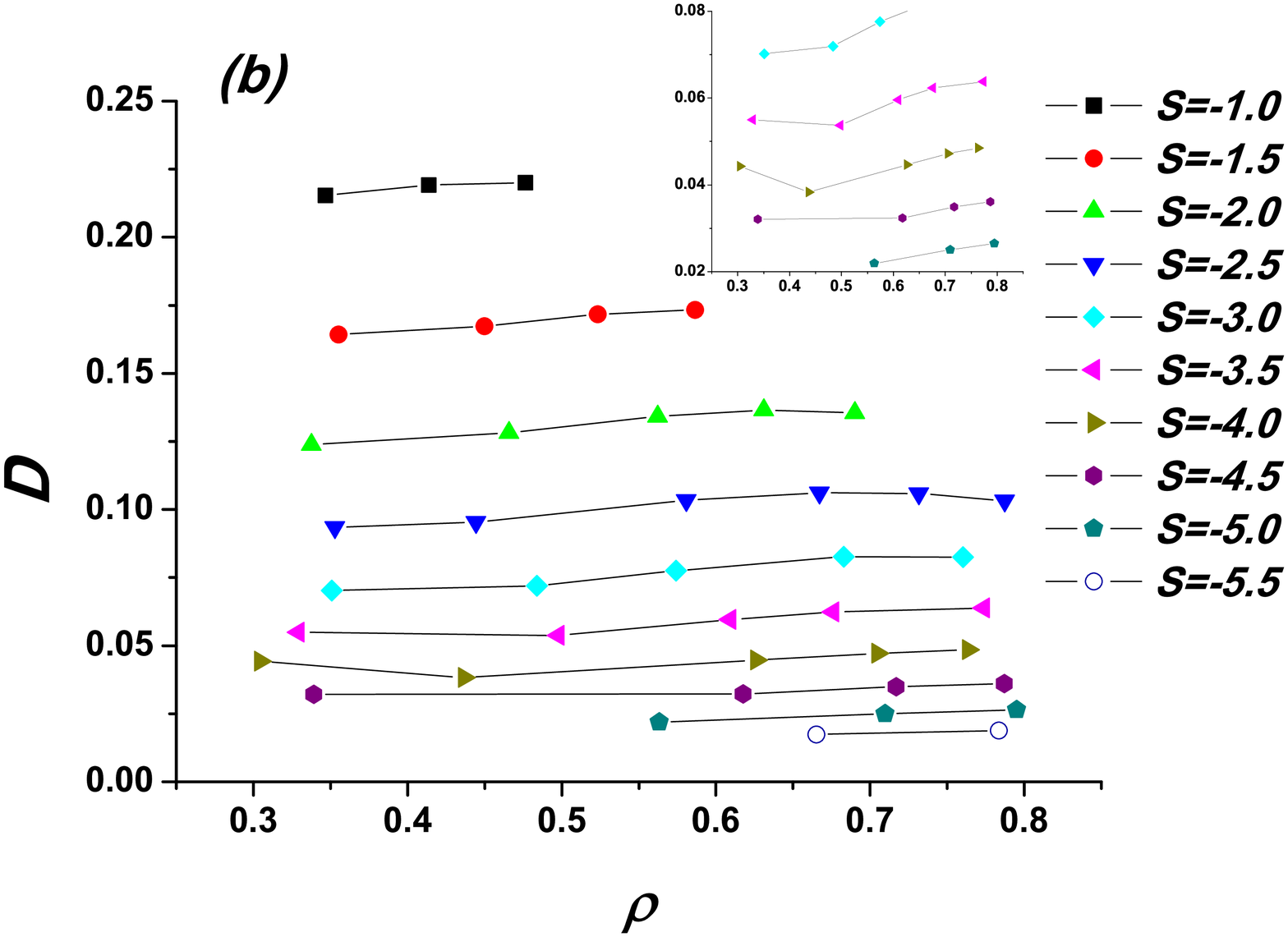}%

\includegraphics[width=7cm, height=7cm]{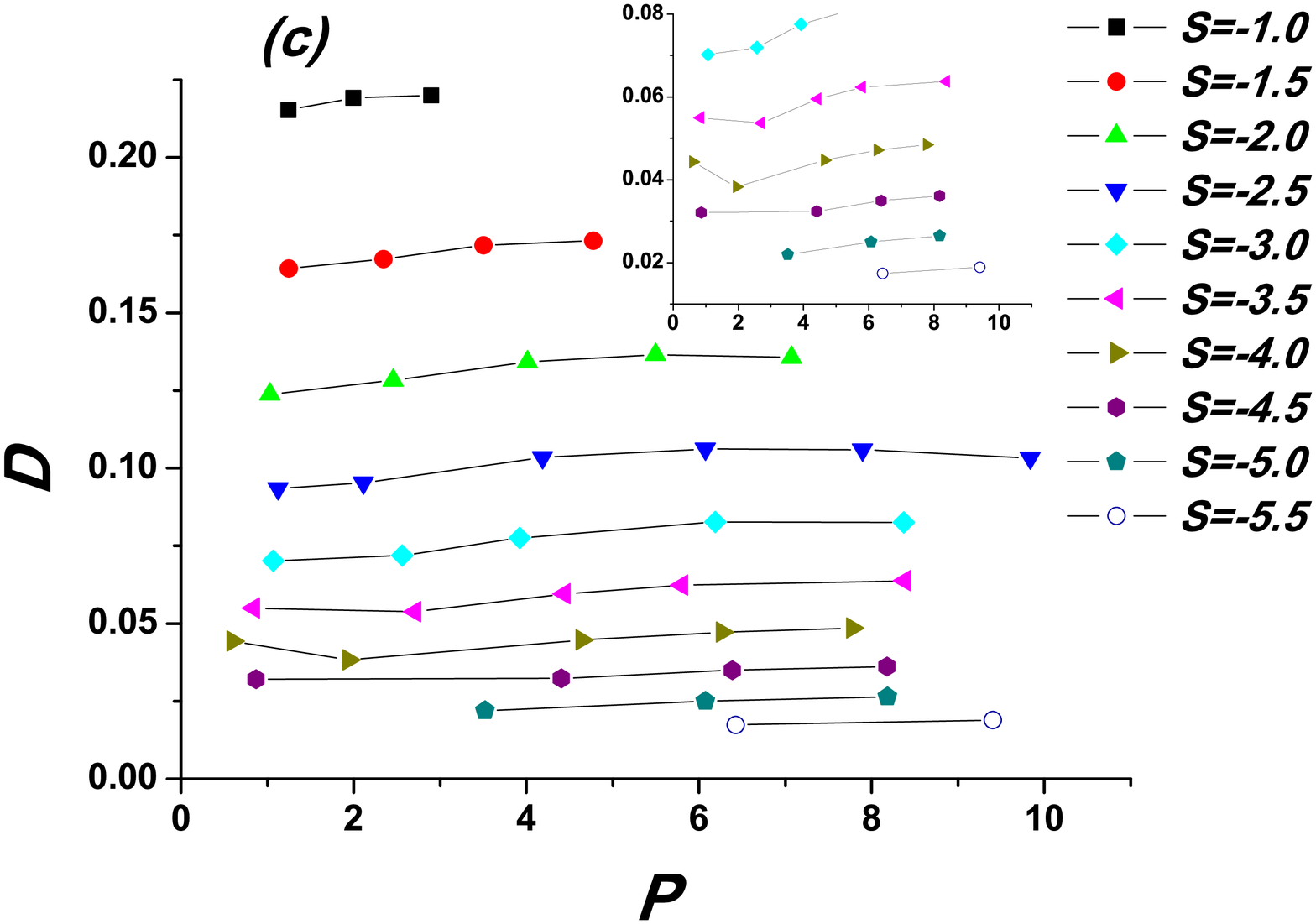}%
\caption{\label{fig:fig3} (Color online). Diffusion coefficient of
the RSS system along adiabats as a function of (a) temperature,
(b) density, and (c) pressure. The insets show anomalous regions
in the corresponding coordinates.}
\end{figure}

The behavior of entropy itself will be discussed below. Here we
give the diffusion coefficients along the adiabats
(Fig.~\ref{fig:fig3} (a)-(c)).

One can see from the Fig.~\ref{fig:fig3} (a)-(c) that the anomaly
takes place along adiabats in all possible coordinates (look, for
example, $S=-4.0$ adiabat). It means that in case of adiabatic
trajectory one can identify the anomalous region monitoring any of
three thermodynamic variables ($P,\rho,T$). However, this
trajectory is rather difficult to realize in simulation or
experiment.

\subsection{Density Anomaly}

\begin{figure}
\includegraphics[width=7cm, height=7cm]{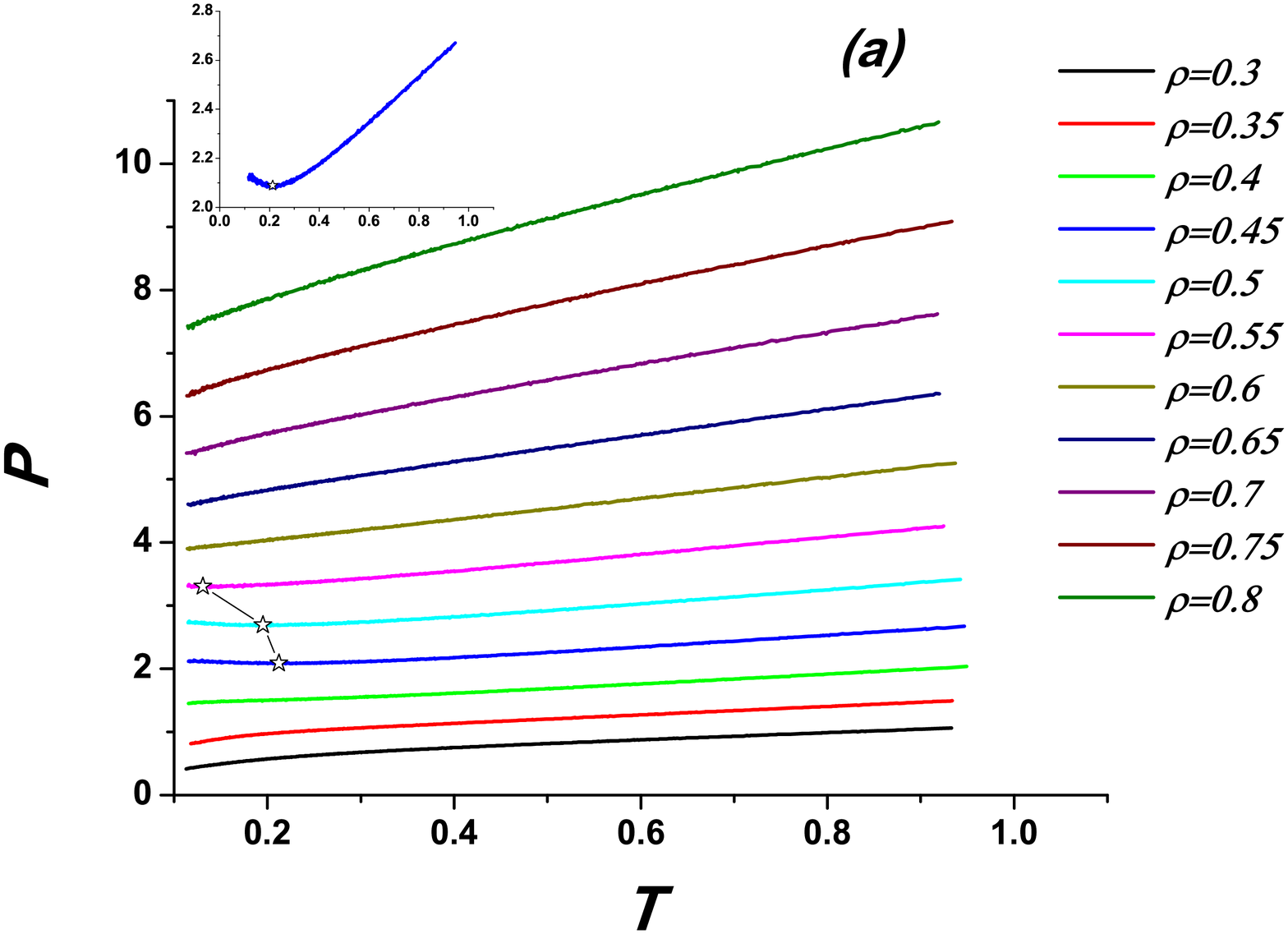}%

\includegraphics[width=7cm, height=7cm]{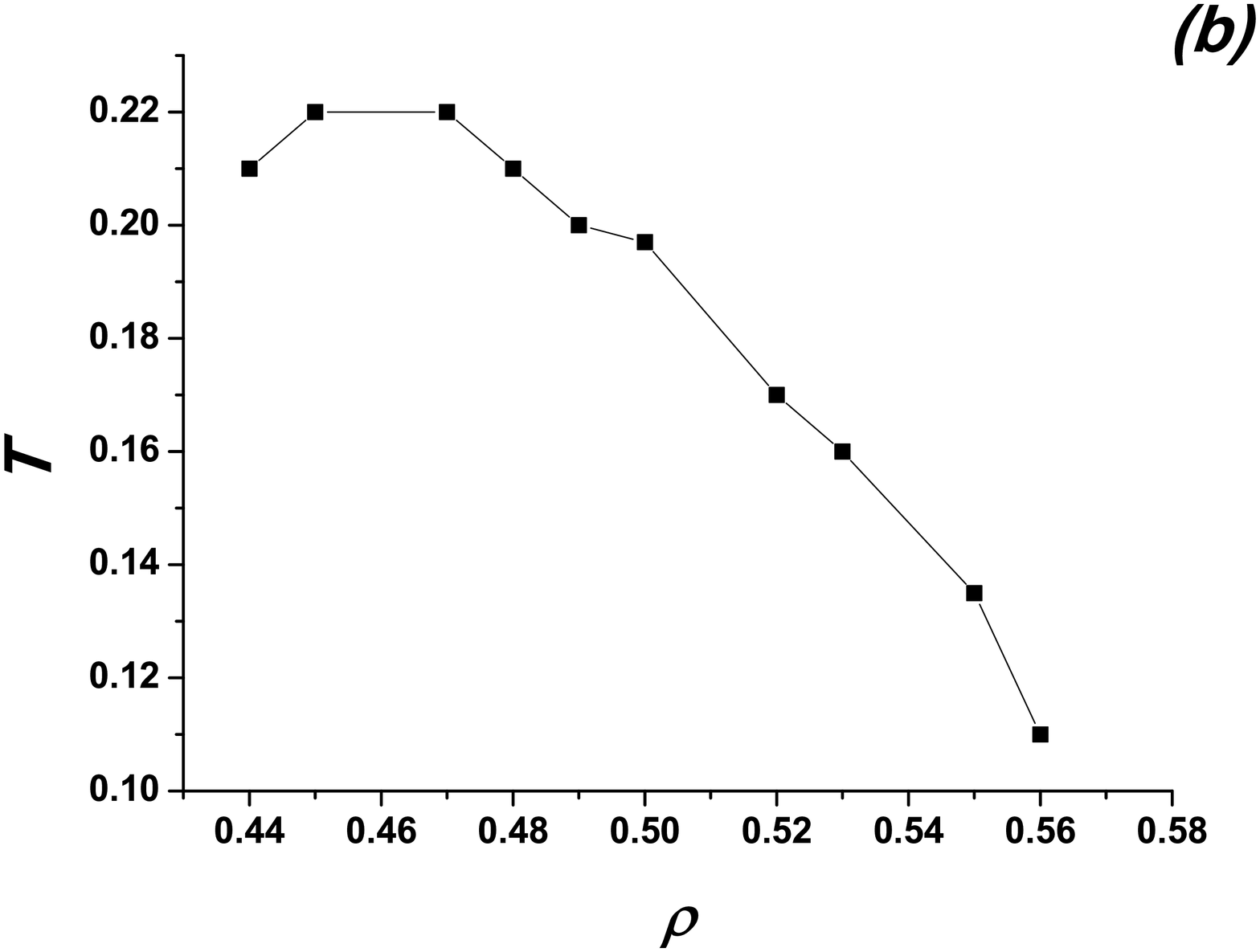}%

\caption{\label{fig:fig4} (Color online). (a) A set of isochors of
RSS. The stars show the location of minimum. The inset enlarges
the $\rho=0.45$ isochor. (b) The location of the minima on
isochors in $\rho - T$ plane.}
\end{figure}

As we mentioned above density anomaly corresponds to appearance of
a minima on isochors of the system. The isochors are shown in the
Fig.~\ref{fig:fig4}(a). It is evident from the figure that some of
the isochors do demonstrate minima. The location of the minimum in
the $\rho-T$ plane is shown in Fig.~\ref{fig:fig4}(b). One can see
from this figure that the region of the density anomaly is located
inside the region of the diffusion anomaly which is consistent
with the general picture corresponding to water.

\begin{figure}
\includegraphics[width=7cm, height=7cm]{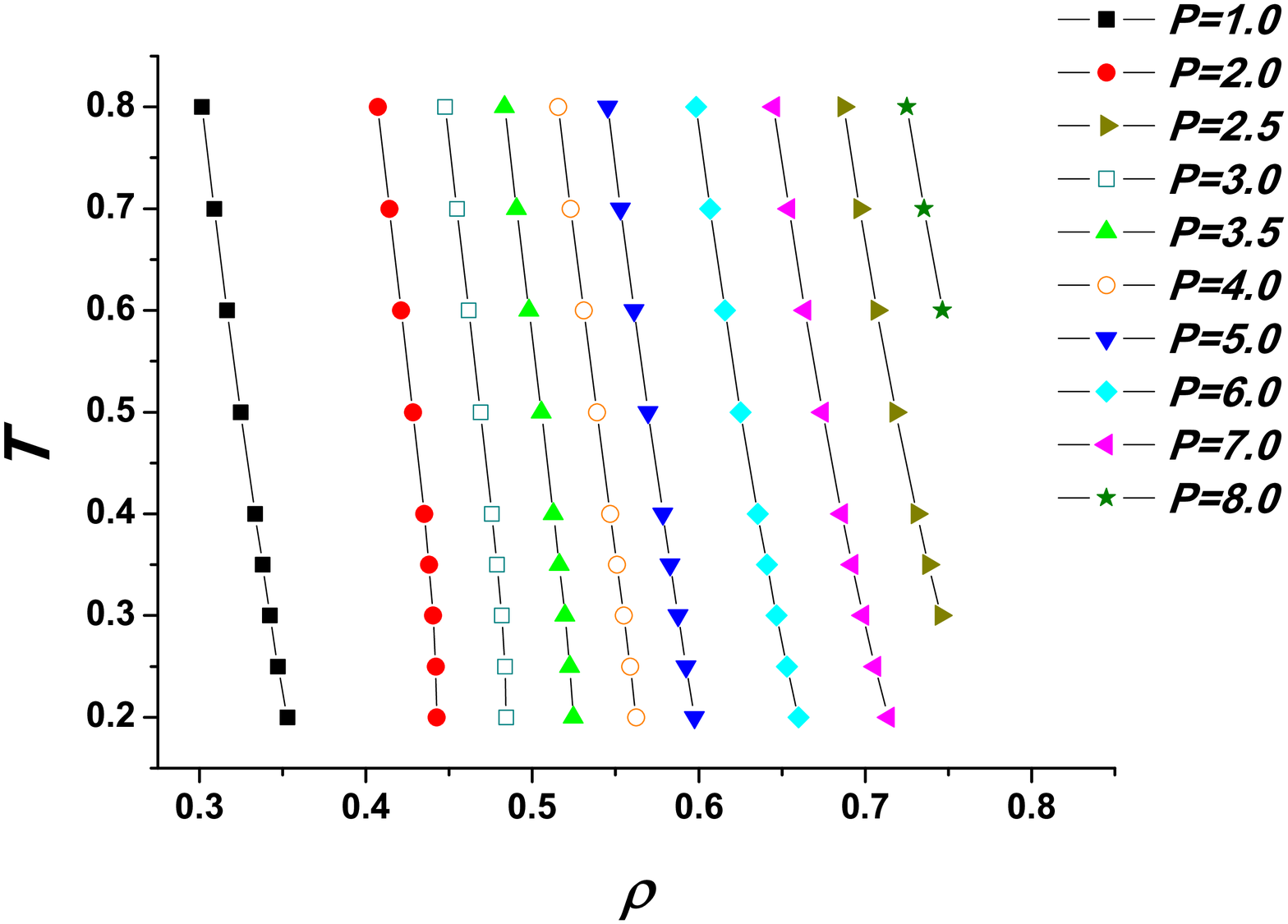}%

\caption{\label{fig:fig5} (Color online). A set of isobars of RSS.
}
\end{figure}

If we turn to the shape of the isobars themselves, that is the
dependence of temperature on density at fixed pressure we do not
find any traces of anomalies there (Fig. ~\ref{fig:fig5}). Like in
the case of diffusion, the curves have negative slope which
approaches zero at low temperatures and densities corresponding to
anomalous regime. However, the curve remains monotonous which
means that there is no sign of density anomaly along isobars.

\begin{figure}
\includegraphics[width=7cm, height=7cm]{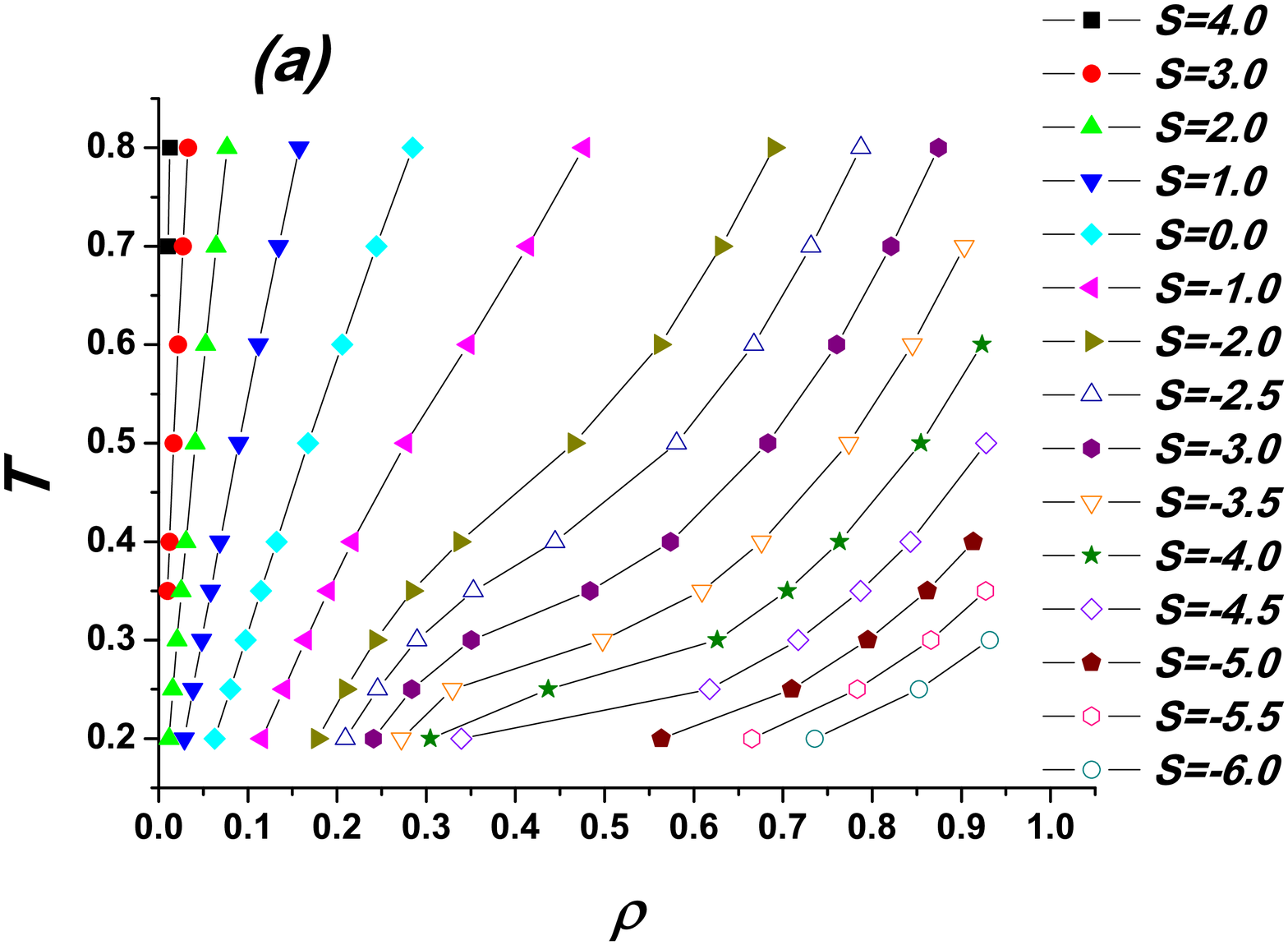}%

\includegraphics[width=7cm, height=7cm]{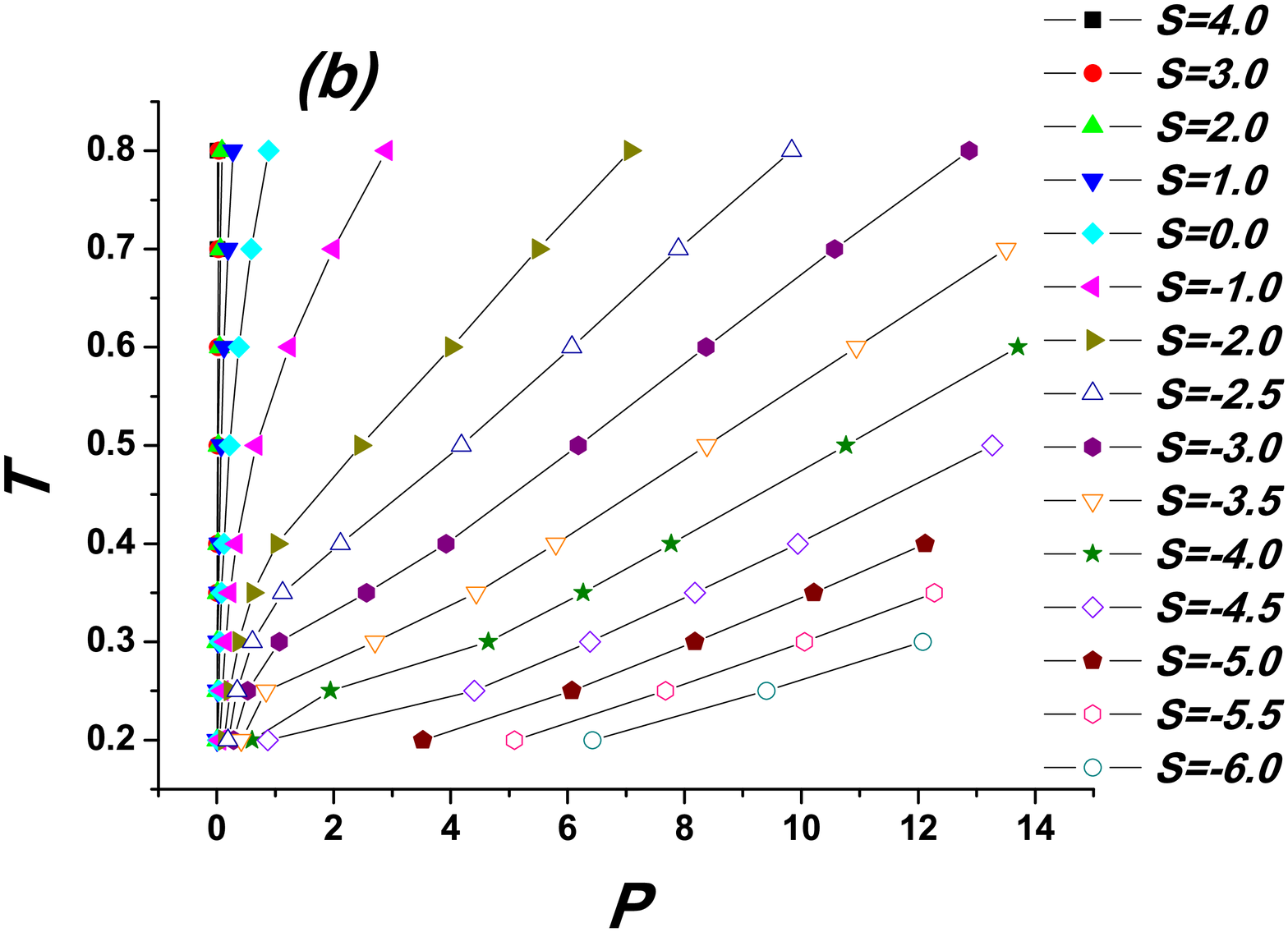}%

\caption{\label{fig:fig6} (Color online). Adiabats of the RSS in
(a) density - temperature and (b) pressure - temperature
coordinates.}
\end{figure}

Figs. ~\ref{fig:fig6}(a) and (b) show adiabats of the RSS in
$\rho-T$ and $P-T$ coordinates. Interestingly, the curves are
again monotonic, but the slope of the curves at low and high
densities (pressures) is very different. At the same time the
middle density (pressure) curves demonstrate a continues change
from the high slope (low-density regime) to low slope
(high-density) regime. It allows to conclude that even if adiabats
show monotonous behavior one can easily identify the location of
anomalous region by the slope change of the curves.

\subsection{Structural Anomaly}

Structural anomaly region can be bounded by using the local order
parameters or by excess entropy minimum and maximum. Here we apply
the definition via excess entropy.

The behavior of excess entropy is qualitatively analogous to the
behavior of diffusion coefficient. Because of this we briefly
describe it here noting that most of the conclusions about
diffusion coefficient along different trajectories can be applied
to excess entropy as well.

\begin{figure}
\includegraphics[width=7cm, height=7cm]{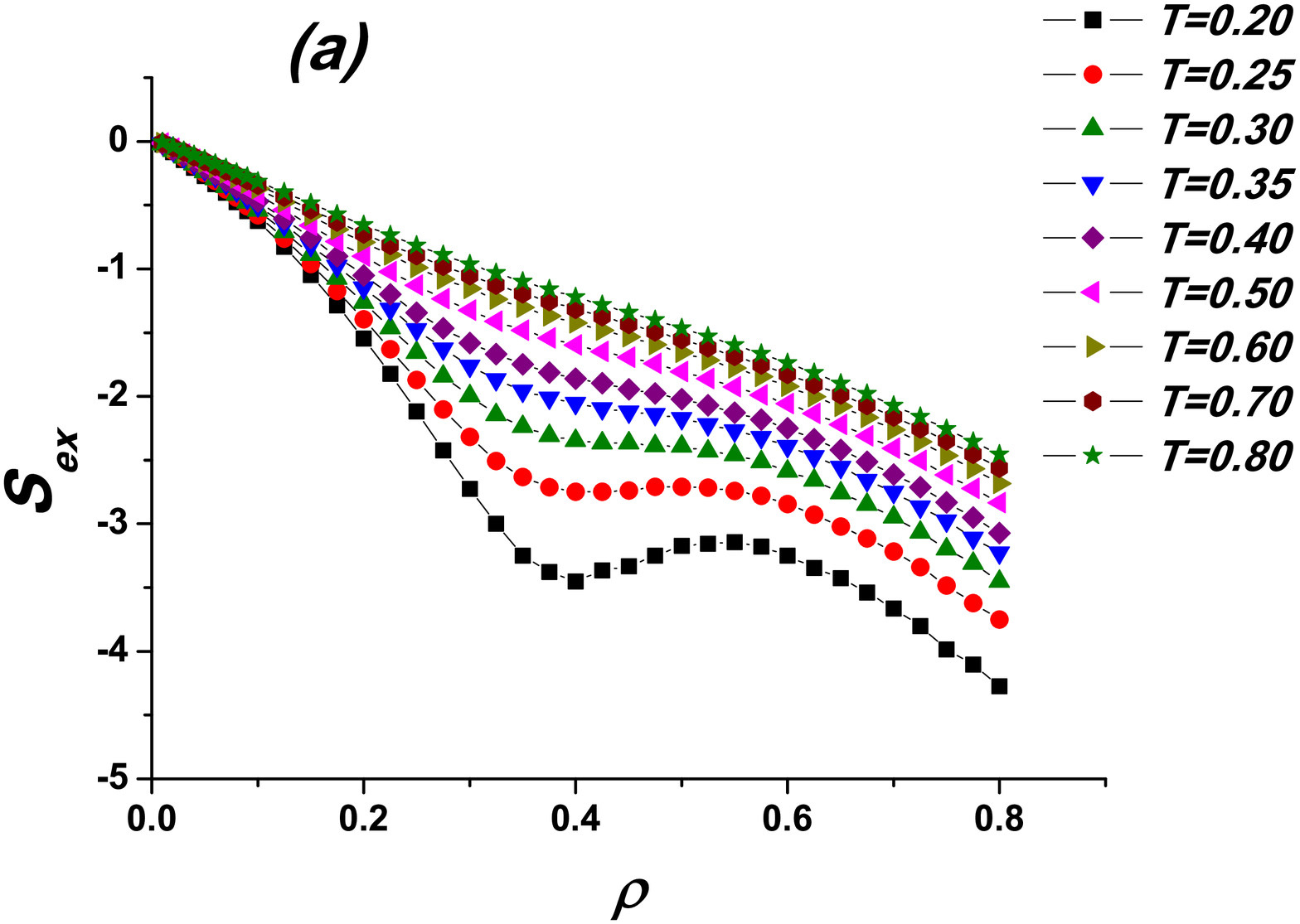}%

\includegraphics[width=5cm, height=7cm]{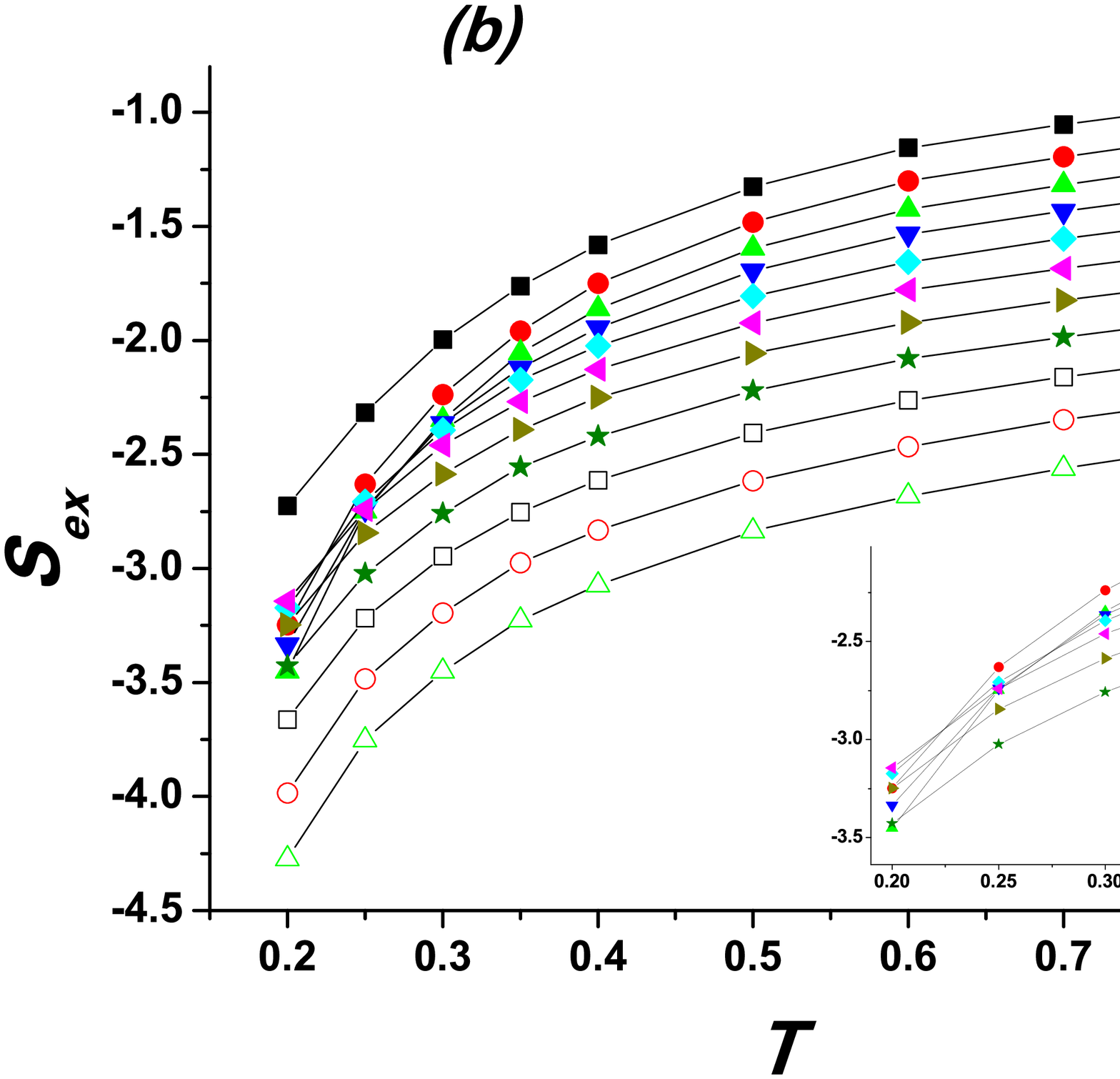}%

\caption{\label{fig:fig7} (Color online). Excess entropy of RSS
fluid along (a) isotherms and (b) isochors.}
\end{figure}

Figs.~\ref{fig:fig7}(a) and (b) show the excess entropy along
isotherms and isochors. Like for the diffusion coefficient, excess
entropy demonstrates anomalous grows in some density range at low
temperatures. At the same time excess entropy is monotonous along
isochors. However, the curves for different isochors cross which
indicates the presence of anomaly.

\begin{figure}
\includegraphics[width=7cm, height=7cm]{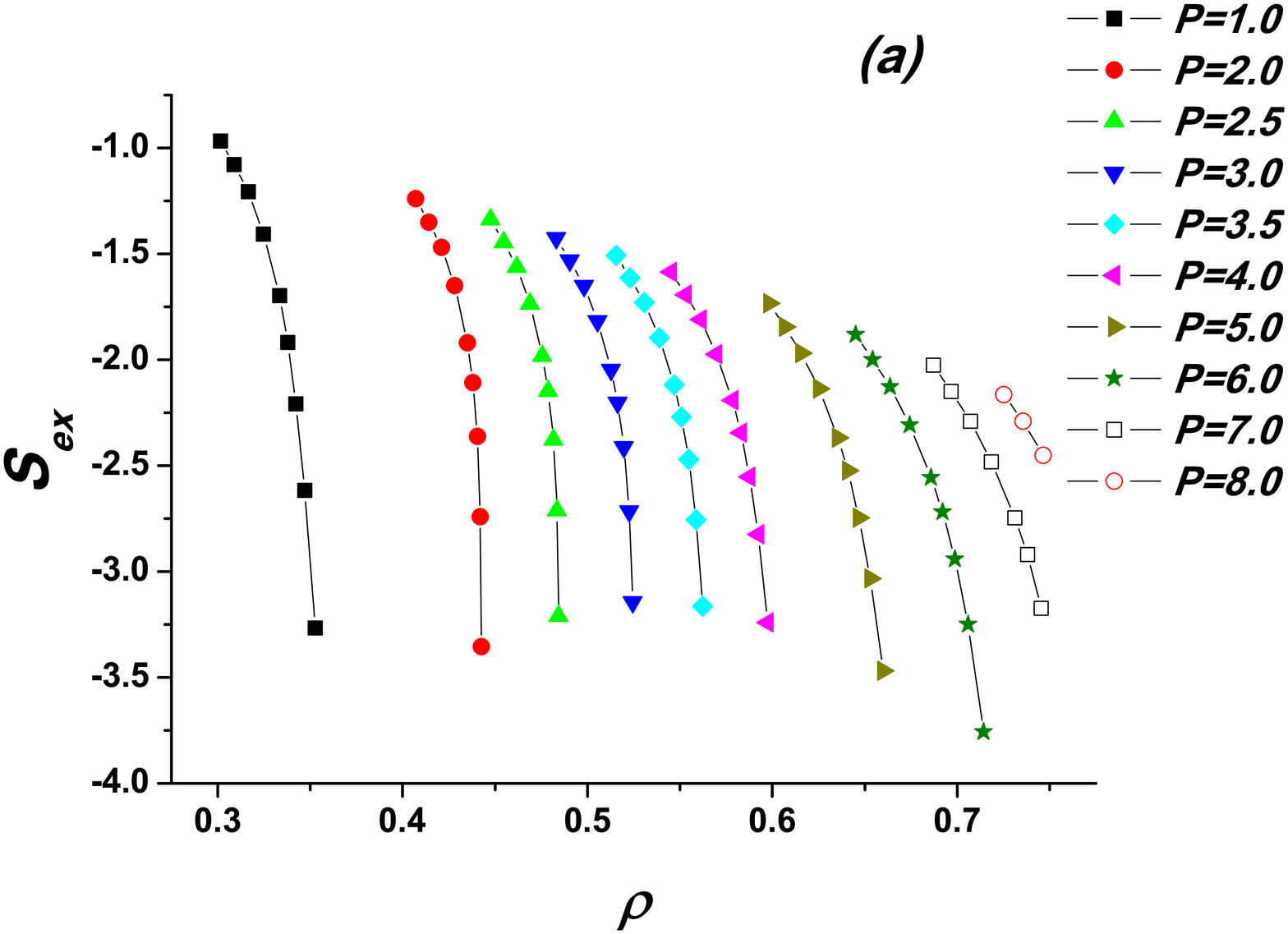}%

\includegraphics[width=7cm, height=7cm]{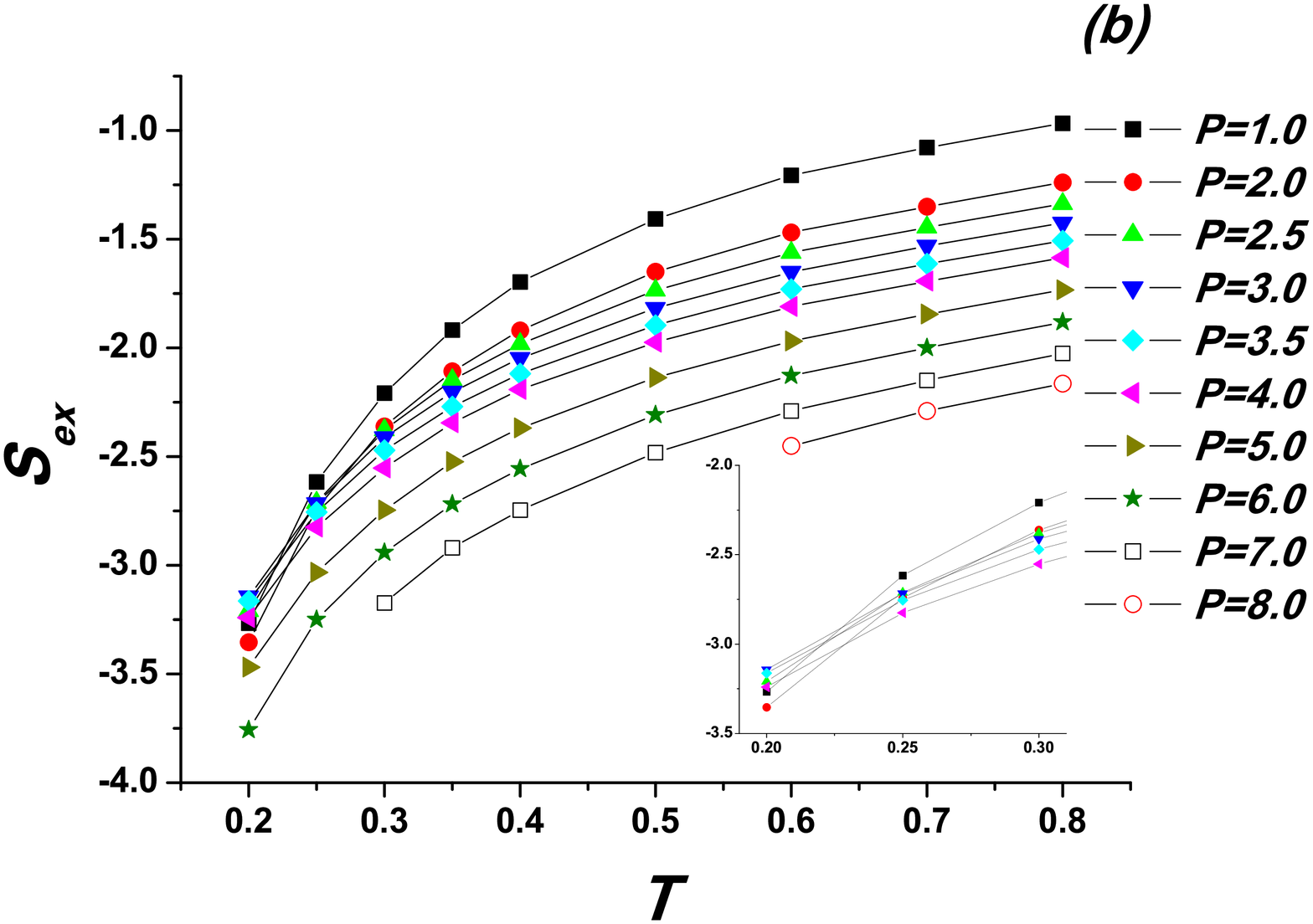}%

\caption{\label{fig:fig8} (Color online). Excess entropy of RSS
fluid along isobars as a function of (a) density and (b)
temperature. The inset in (b) shows the cross of the curves at low
temperatures.}
\end{figure}

The excess entropy along isobars is monotonically decreasing
function of density and monotonically increasing function of
temperature. However, the curves cross at low temperatures
indicating the presence of anomalies as it was discussed for the
case of diffusion.

It is important to note that the range of densities of structural
anomalies is wider then the one of diffusion anomaly which is
consistent with the literature data for core-softened systems.

To summarize this section, we emphasize that the anomalous
behavior can be seen only along some particular trajectories in
$(P,\rho,T)$ space. For example, diffusion and structural
anomalies are visible along isotherms while density anomaly -
along isochors. Other meaningful trajectories (isobars and
adiabats) can identify the presence of anomalies via crossing of
the curves, but they do not allow easily identify the boundaries
of the anomalous region.

\section{IV. Rosenfeld scaling}

In 1977 Rosenfeld proposed a connection between thermodynamic and
dynamical properties of liquids \cite{ros1,ros2}. The main
Rosenfeld's statement claims that the transport coefficients are
exponential functions of the excess entropy. In order to write the
exponential relations Rosenfeld introduced reduction of the
transport coefficients by some macroscopic parameters of the
system. For the case of diffusion coefficient one writes: $D^*=D
\frac{\rho ^{1/3}}{(k_BT/m)^{1/2}}$, where $m$ is the mass of the
particles. The Rosenfeld scaling rule can be written as:

\begin{equation}
 D^*=A \cdot e^{BS_{ex}},
\end{equation}
where $A$ and $B$ are constants.

In his original works Rosenfeld considered hard spheres, soft
spheres, Lennard-Jones system and one-component plasma
\cite{ros1,ros2}. After that the excess entropy scaling was
applied to many different systems including core-softened liquids
\cite{errington,errington2,india1,indiabarb,weros}, liquid metals
\cite{liqmet1,liqmet2}, binary mixtures \cite{binary1,binary2},
ionic liquids \cite{india2,ionicmelts}, network-forming liquids
\cite{india1,india2}, water \cite{buldwater}, chain fluids
\cite{chainfluids} and bounded potentials
\cite{weros,klekelberg,klekelberg1}.

In our recent publication \cite{weros,werostr} we showed that for
the case of the core-softened fluids the applicability of
Rosenfeld relation depends on the trajectory. In particular,
Rosenfeld relation is applicable along isochors, but it is not
applicable along isotherms.

The breakdown of the Rosenfeld relation along isotherms can be
seen from the following speculation. The regions of different
anomalies do not coincide with each other. In particular, in the
case of core-softened fluids the diffusion anomaly region is
located inside the structural anomaly one. It means that there are
some regions where the diffusion is still normal while the excess
entropy is already anomalous. But this kind of behavior can not be
consistent with the Rosenfeld scaling law.

\begin{figure}
\includegraphics[width=7cm, height=7cm]{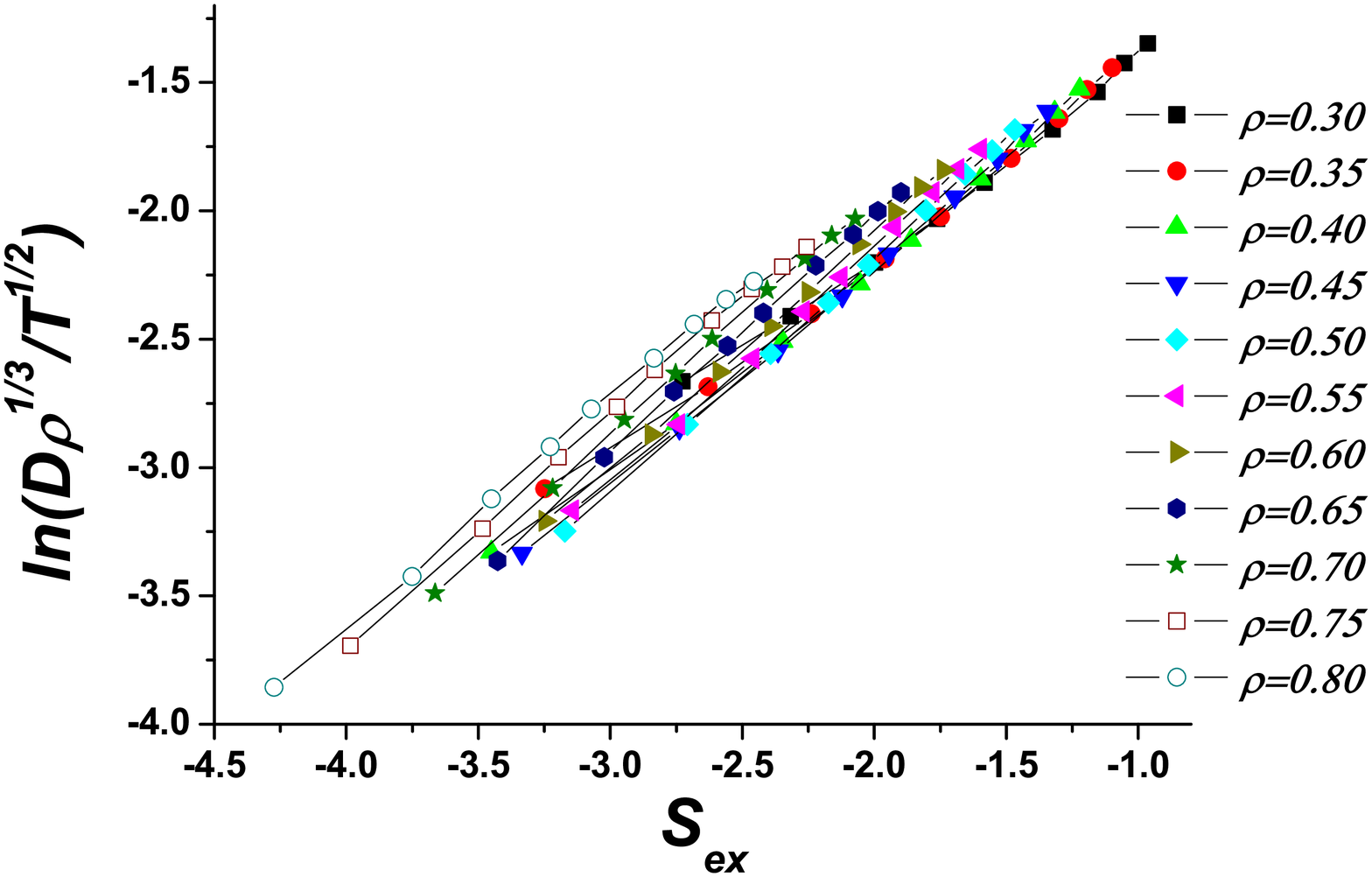}%

\includegraphics[width=7cm, height=7cm]{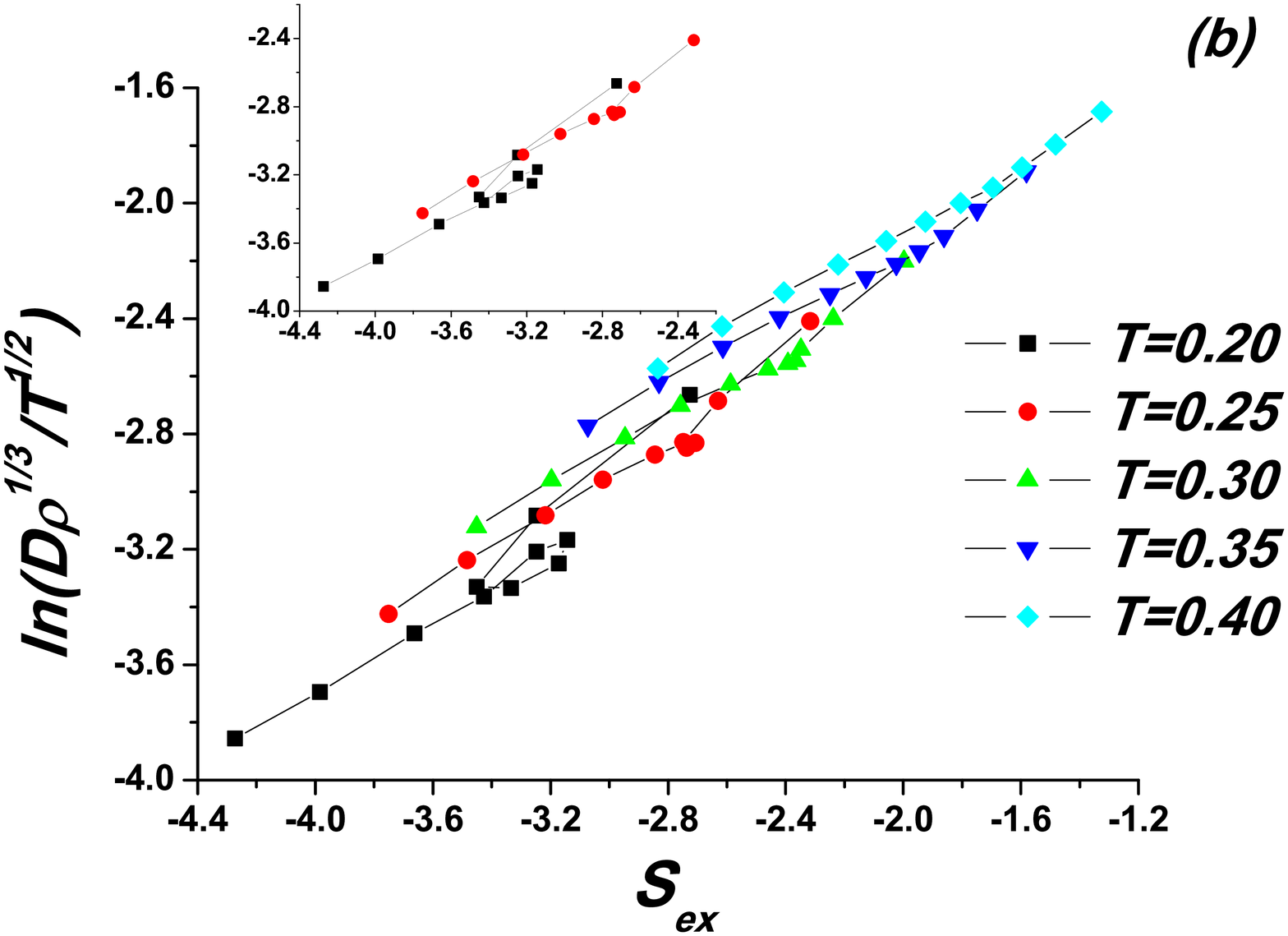}%

\includegraphics[width=7cm, height=7cm]{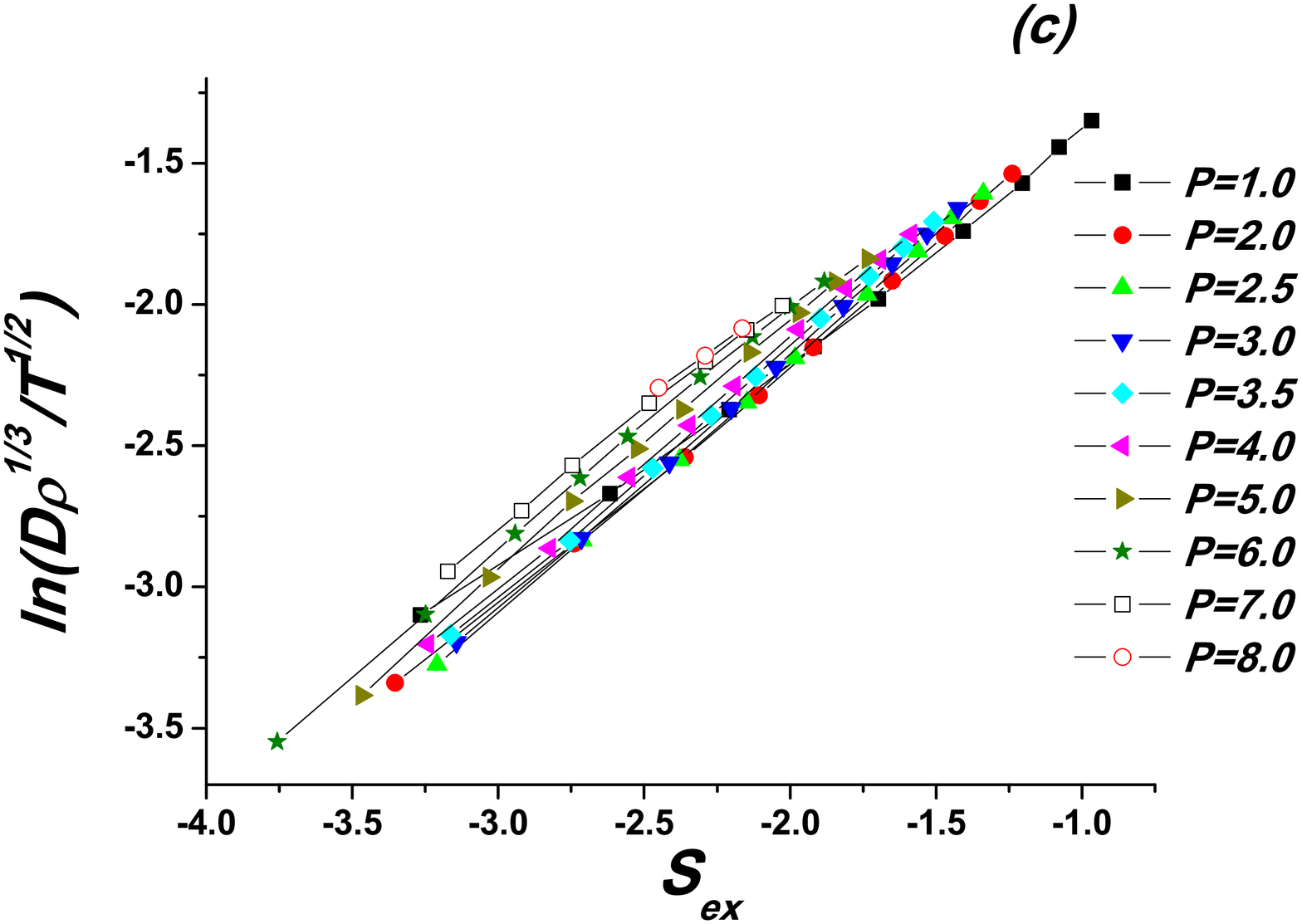}%

\caption{\label{fig:fig9} (Color online). Rosenfeld relation for
RSS along (a) isochors, (b) isotherms, and (c) isobars.}
\end{figure}

However, from this speculation it follows that the Rosenfeld
scaling should hold true along the trajectories which do not
contain anomalies, i.e. isochors and isobars. In our recent
publication \cite{werostr} we considered the Rosenfeld relation
along isotherms and isochors. Here we bring these trajectories for
the sake of completeness and add the verification of the Rosenfeld
relation along isobars (Fig. ~\ref{fig:fig9}(a) - (c)). One can
see that the Rosenfeld relation does break down along isotherms
which is consistent with the speculation above. At the same time
it holds true along both isochors and isobars which is consistent
with the monotonous behavior of both diffusion coefficient and
excess entropy along these trajectories.

This observation makes evident that Rosenfeld relation is valid
only along the trajectories without anomalous behavior.

\section{V. Conclusions}

In conclusion, in the present article we carry out a molecular
dynamics study of the core-softened system (RSS) and show that the
existence of the water-like anomalies in this system depends on
the trajectory in $P-\rho-T$ space along which the behavior of the
system is studied. For example, diffusion and structural anomalies
are visible along isotherms, but disappears along the isochores
and isobars, while density anomaly exists along isochors.  We
analyze the applicability of the Rosenfeld entropy scaling
relations to this system in the regions with the water-like
anomalies. It is shown that the validity of the of Rosenfeld
scaling relation for the diffusion coefficient also depends on the
trajectory in the $P-\rho-T$ space along which the kinetic
coefficients and the excess entropy are calculated. In particular,
it is valid along isochors and isobars, but it breaks down along
isotherms. The breakdown of the Rosenfeld relation along isotherms
can be explained in the following way. The regions of different
anomalies do not coincide with each other. In particular, in the
case of core-softened fluids the diffusion anomaly region is
located inside the structural anomaly one. It means that there are
some regions where the diffusion is still normal while the excess
entropy is already anomalous. But this kind of behavior can not be
consistent with the Rosenfeld scaling law.

\bigskip

\begin{acknowledgments}
We thank V. V. Brazhkin for stimulating discussions. Y.F. also
thanks the Russian Scientific Center Kurchatov Institute and Joint
Supercomputing Center of Russian Academy of Science for
computational facilities. The work was supported in part by the
Russian Foundation for Basic Research (Grant No 08-02-00781).
\end{acknowledgments}


\end{document}